\documentclass[apjl, iop]{emulateapj}

\newcommand  \mgii  {\ifmmode {\rm Mg}{\textsc{ii}} \else Mg\,{\sc ii}\fi}
\newcommand  \MGII  {\ifmmode {\rm Mg}\,{\sc ii}\,\lambda2798 \else Mg\,{\sc ii}\,$\lambda2798$\fi}
\newcommand  \siiv  {\ifmmode {\rm Si}\, {\sc iv}\ \else Si\,{\sc iv}\fi}
\newcommand  \SIIV  {\ifmmode {\rm Si}\,{\sc iv}\,\lambda1399 \else Si\,{\sc iv}\,$\lambda1399$\fi}
\newcommand  \cv  {\ifmmode {\rm C}\, {\sc v}\ \else C\,{\sc v}\fi}
\newcommand  \aliii  {\ifmmode {\rm Al}{\textsc{iii}} \else Al\,{\sc iii}\fi}
\newcommand  \civ  {\ifmmode {\rm C}\, {\sc iv}\ \else C\,{\sc iv}\fi}
\newcommand  \ciii  {\ifmmode {\rm C}\, {\sc iii}\ \else C\,{\sc iii}\fi}
\newcommand  \CIV  {\ifmmode {\rm C}\,{\sc iv}\,\lambda1549 \else C\,{\sc iv}\,$\lambda1549$\fi}
\newcommand  \NV  {\ifmmode {\rm N}\,{\sc v}\,\lambda1240 \else N\,{\sc v}\,$\lambda1240$\fi}
\newcommand  \nv  {\ifmmode {\rm N}\,{\sc v}\ \else N\,{\sc v}\fi}
\newcommand  \ovi    {\ifmmode \left[{\rm O}\,{\textsc vi}\right] \else O\,{\sc vi}\fi}
\newcommand  \LyA  {\ifmmode {\rm Lyman}\,{\sc $\alpha$}\,\lambda1216 \else Lyman\,{\sc $\alpha$}\,$\lambda1216$\fi}
\newcommand  \lya {\ifmmode {\rm Lyman}\,{\sc $\alpha$}\ \else Lyman\,{\sc $\alpha$}\fi}
\newcommand  \feii     {Fe\,{\sc ii}}

\newcommand  \ALIII  {\ifmmode {\rm Al}\,{\sc iii}\,\lambda1857 \else Al\,{\sc iii}\,$\lambda1857$\fi}

\newcommand{\kms}{\ifmmode {\rm km\,s}^{-1} \else km\,s$^{-1}$ \fi}

\usepackage{ulem}

\shorttitle{BAL disappearance/emergence in a WLQ}
\shortauthors{Yi et al.}

\begin{document}


\title{ Broad absorption line disappearance/emergence in multiple ions in a weak emission-line quasar} 

\author{W. Yi\altaffilmark{1,2,3}, M. Vivek\altaffilmark{1}, W.~N. Brandt\altaffilmark{1,4,5},  T. Wang\altaffilmark{6}, J. Timlin\altaffilmark{1}, N. Filiz Ak\altaffilmark{7,8},  D.~P. Schneider\altaffilmark{1,4}, J.~P.~U. Fynbo\altaffilmark{9}, Q. Ni\altaffilmark{1}, F. Vito\altaffilmark{10,11}, B.~L.  Indahl\altaffilmark{12}, Sameer\altaffilmark{1}   }


\altaffiltext{1}{Department of Astronomy \& Astrophysics, The Pennsylvania State University, 525 Davey Lab, University Park, PA 16802, USA}  
\altaffiltext{2}{Yunnan Observatories, Kunming, 650216, China}
\altaffiltext{3}{Key Laboratory for the Structure and Evolution of Celestial Objects, Chinese Academy of Sciences, Kunming 650216, China} 
\altaffiltext{4}{Institute for Gravitation and the Cosmos, The Pennsylvania State University, University Park, PA 16802, USA}
\altaffiltext{5}{Department of Physics, The Pennsylvania State University, University Park, PA 16802, USA}
\altaffiltext{6}{CAS Key Laboratory for Research in Galaxies and Cosmology, Department of Astronomy, University of Science and Technology of China, China}
\altaffiltext{7}{Faculty of Sciences, Department of Astronomy and Space Sciences, Erciyes University, 38039, Kayseri, Turkey}
\altaffiltext{8}{Astronomy and Space Sciences Observatory and Research Center, Erciyes University, 38039, Kayseri, Turkey} 
\altaffiltext{9}{Cosmic Dawn Center (DAWN), Niels Bohr Institute, University of Copenhagen, Juliane Maries Vej 30, DK-2100 Copenhagen $\O$; DTU-Space, Technical University of Denmark,  Elektrovej 327, DK-2800 Kgs. Lyngby, Denmark}  
\altaffiltext{10}{Instituto de Astrofísica and Centro de Astroingeniería, Facultad de Física, Pontificia Universidad Catolica de Chile, Casilla 306, Santiago 22, Chile}
\altaffiltext{11}{Chinese Academy of Sciences South America Center for Astronomy, National Astronomical Observatories, CAS, Beijing 100012, China}
\altaffiltext{12}{Department of Astronomy, University of Texas, Austin, TX 78712, USA}

\begin{abstract}
We report the discovery of disappearance of \mgii, \aliii, \civ, and \siiv\ broad absorption lines (BALs) at the same velocity (0.07c), accompanied by a new \civ\ BAL emerging at a higher velocity (up to 0.11c), in the quasar  J0827+4252 at $z=2.038$. This is the first report of BAL disappearance (i) over \mgii, \aliii, \civ, and \siiv\ ions and (ii) in a weak emission-line quasar (WLQ). The discovery is based on four spectra from the SDSS and one  follow-up spectrum from HET/LRS2. 
The simultaneous \civ\ BAL disappearance and emergence at different velocities, together with no variations in the CRTS light curve, indicate that ionization changes in the absorbing material are unlikely to cause the observed BAL variability. 
Our analyses reveal that transverse motion is the most likely dominant driver of the BAL disappearance/emergence. 
Given the presence of mildly relativistic BAL outflows and an apparently large \civ\ emission-line blueshift that is likely associated with strong bulk outflows in this WLQ, J0827+4252 provides a notable opportunity to study extreme quasar winds and their potential in expelling material from inner to large-scale regions.
\end{abstract}


\section{Introduction}

Broad absorption line quasars (BALQSOs; \citealp{Weymann91}) make up $\approx$ 15\% of quasars discovered to date (e.g., \citealp{Trump06,Gibson09}), but their intrinsic fraction may be up to $\approx$ 40\% due to selection effects (e.g., \citealp{Allen:2011}). They are often classified into high-ionization BALQSOs  (HiBALs) and low-ionization BALQSOs (LoBALs). 
HiBAL quasars show absorption features only from relatively high-ionization species such as \civ\ and \siiv. LoBAL quasars possess high-ionization features plus absorption from low-ionization species, typically  \aliii\ and \mgii. 
Almost all BALQSOs are characterized by blueshifted absorption troughs imprinted on their spectra, signaling outflows along our line of sight (LOS). 

The inner circumnuclear regions of quasars cannot be spatially resolved with current technology, and thus we cannot directly observe the physical processes that cause BAL outflows. However, BAL variability provides one of the most-powerful diagnostics for exploring the nature and origin of such outflows. Statistically, BAL variability is widely explained by transverse motions of BAL absorbers across our LOS or ionization changes in response to continuum variations  (e.g., \citealp{Cap11,Filizak13,Wang15}). 

BAL emergence or disappearance, an extreme and relatively rare form of BAL variability, is often interpreted as gas moving into or out of our LOS (e.g., \citealp{Junkkarinen2001,Hamann08,Leighly09,Krongold10,Hall2011,Vivek2012,Rafiee2016}).  Some studies, however, report observational  evidence in support of ionization changes (e.g., \citealp{McGraw2017,Stern2017,Vivek2018}). 
A few studies based on relatively large samples confirm that both the ionization-change and transverse-motion mechanisms account for most BAL-emergence/disappearance phenomena (e.g., \citealp{Filizak12,Wang15,McGraw2017,DeCicco18,Rogerson18,Sameer2019}). 
The dominant driver of BAL emergence/disappearance in individual BALQSOs, however, often remains uncertain due to a lack of sufficient observational evidence to break degeneracies inherent in BAL variability. In fact, BAL disappearance or emergence phenomena have been reported mostly for weak BAL troughs caused by a single ionized species. Only a few investigations have reported BAL emergence or disappearance occurring over multiple-ion BAL troughs (e.g., \citealp{Hamann08,Filizak12,McGraw2017,DeCicco18,Rogerson18}). 
To date, no BAL-variability studies have reported a case where BAL disappearance occurred over all of  the commonly observed \mgii, \aliii, \civ, and \siiv\ ions at the same velocity.

Weak emission-line quasars (WLQs) are a notable ``extreme'' subset of the quasar population (e.g., \citealp{Fan99,Diamond09,Shemmer10,Wu15,Luo15,Plotkin15}), although previous studies have not reached consensus on the detailed nature of WLQs. Recent WLQ studies, mainly from an X-ray perspective, broadly support a ``shielding'' scenario for WLQs that is associated with high Eddington ratios (e.g., \citealp{Wu12,Luo15,NiQ18}). Likewise, some BALQSO models also require a ``shielding'' mechanism by which high-energy photons are blocked from reaching the UV-absorbing clouds (e.g., \citealp{Proga00}), although a dense outflow could have a sufficiently low ionization level itself without shielding (e.g., \citealp{Hamann13,Baskin2014a}). 
Regardless of specific models, both BALQSOs and WLQs appear to have high-velocity outflows driven by radiation pressure. Indeed, UV BALs often show high LOS velocities ($>$0.03$c$; e.g. \citealp{Hamann08,Vivek2012,Rogerson16}) while both BALQSOs and WLQs tend to have larger \civ\ broad emission-line (BEL) blueshifts than non-BAL quasars (e.g., \citealp{Richards11,Luo15}). 
However, previous WLQ studies were almost entirely based on non-BAL objects. In this context, studying BALs in WLQs is valuable for understanding the connection between the two quasar subsets.  

In this Letter, we report the first discovery of dramatic BAL disappearance over \mgii, \aliii, \civ, and \siiv\  species at the same velocity ($\sim$0.07$c$) followed by a new, stronger \civ\ BAL emerging at a higher velocity (up to 0.11$c$) in a LoBAL WLQ, J082747.14+425241.1 (hereafter J0827+4252) at $z=2.038$. 

\section{Observations and data reduction}

\begin{table}[h]
\centering
 \caption{ Spectroscopic observations of J0827+4252}
 \begin{tabular}{lcccc}
  \hline\noalign{\smallskip}
Instrument & S/N &  Spectral  &  Integration  &  Observation  \\
Name &  &  Coverage  &  Time  &  Date \\
           &  &  (\AA)  &  (seconds)  &  (MJD) \\
  \hline\noalign{\smallskip}
SDSS & 5.1 & 3800--9200 & 9120  & 52266  \\ 
SDSS & 3.4 & 3800--9200 & 3503 & 54524   \\
BOSS & 6.8 & 3650--10400 & 4504 & 55513   \\
BOSS & 6.7 & 3650--10400 & 5400 & 57063   \\
HET/LRS2 & 7.2 & 3700--7000 & 2700 & 58212   \\

  \noalign{\smallskip}\hline
\end{tabular}
\label{table1}
\tablecomments{The S/N values represent the average S/N at $1750<\lambda_{\rm rest}<1800$ \AA. All   spectra have a similar spectral resolution ($R\approx$~1800) at $\lambda_{\rm obs}=4000$ \AA. }
\end{table}

J0827+4252 is a LoBAL quasar selected from our \mgii-BAL quasar sample (Yi et al., in prep) that is dedicated to the large-scale investigation of \mgii-BAL variability using multi-epoch spectra from the Sloan Digital Sky Survey-I/II (hereafter SDSS; \citealp{York00}) and the Baryon Oscillation Spectroscopic Survey of SDSS-III (hereafter BOSS; \citealp{Eisenstein11,Dawson13}). SDSS and BOSS are wide-field, large sky survey projects using the same dedicated 2.5-m telescope at the Apache Point Observatory, New Mexico \citep{Gunn06, Smee13}.  
Spectra for this quasar were taken at four, well-separated epochs (see Table \ref{table1}), and the fully reduced spectra were retrieved directly from the SDSS DR14 archive. 
In our sample, we noticed that  \mgii, \aliii, \civ\ and \siiv\ BALs at the same velocity completely disappeared in this source. Therefore, this quasar was ranked as the highest priority for subsequent spectroscopic observations.

We obtained additional spectra for this quasar using the blue arm of the Low-Resolution Spectrograph-2 (LRS2; \citealp{Chonis14}) mounted on the Hobby-Eberly Telescope (HET; \citealp{Ramsey98}) on April 4, 2018. 
The newly obtained spectroscopic data were processed with the LRS2 pipeline (\citealp{Davis18}; Indahl et al. in prep), which includes standard long-slit spectral extraction procedures, but does not yet include flux calibration. However, the lack of flux calibration for the HET/LRS2 spectra does not affect the identification and measurement of BAL features. 
To calibrate the flux of the HET/LRS2 spectrum, we first fit a low-order polynomial to the spectrum and then scaled the spectrum so that the polynomial fit matches the reddened power-law fit (see Section \ref{con_temp_fit}) for the BOSS spectrum at MJD = 55513. In this process, we assumed that the HET/LRS2 spectrum and the BOSS spectrum at MJD = 55513 have the same continuum flux.

For better visual clarity, we smooth all the raw spectra with a 25-pixel Savitzky-Golay filter window as shown in Figure~\ref{J0827all_spec_fit}. This smoothing does not affect the appearance of the broad emission/absorption lines.

\section{Observational results} \label{con_temp_fit}

\begin{figure*}
\center{}
  \includegraphics[height=16cm,width=18cm,  angle=0]{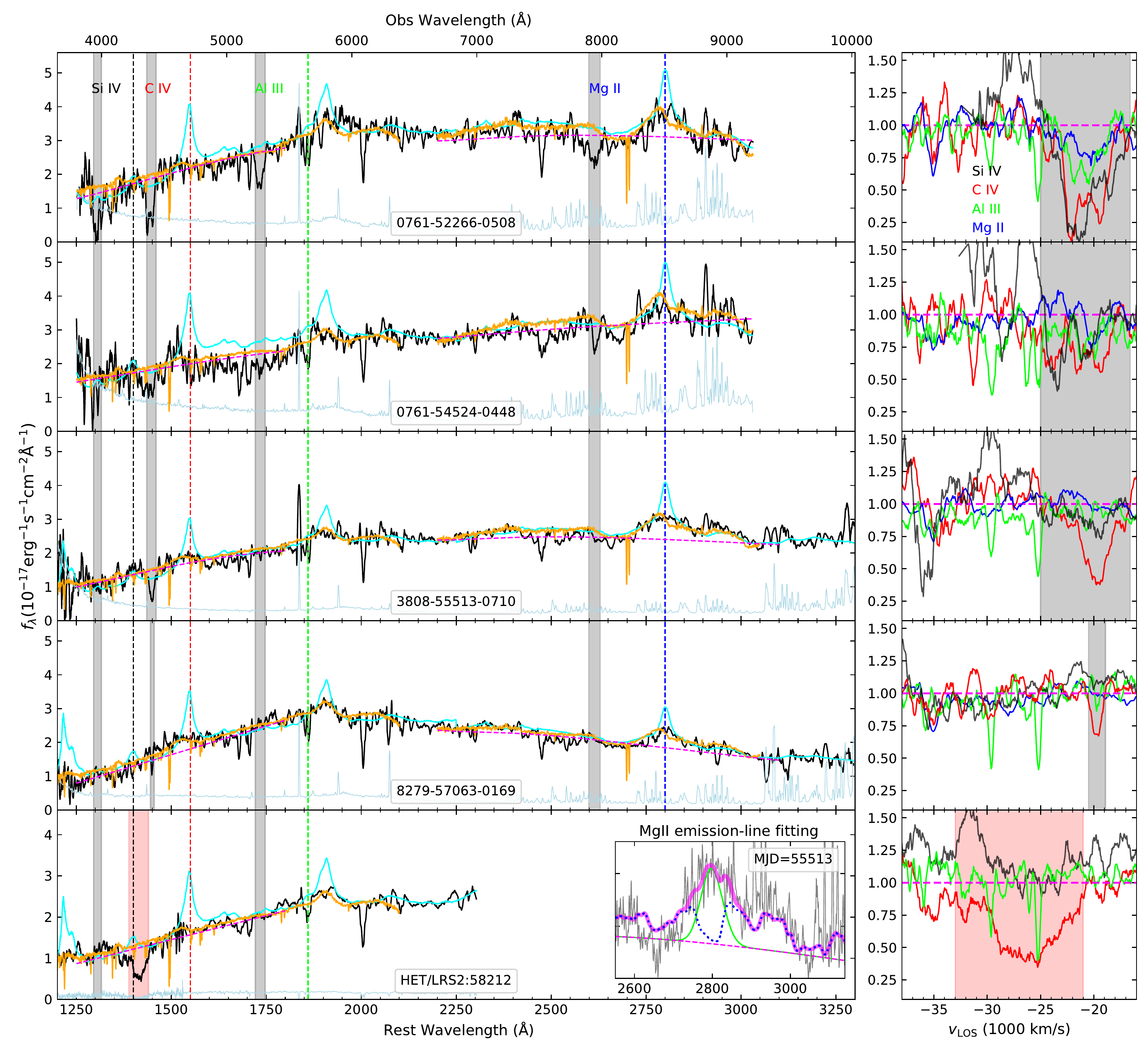} 
      \caption{Left panels: different-epoch spectra with corresponding continuum (dashed magenta) and template fits using the composite template (cyan; \citealt{VandenBerk01}) and the WLQ spectrum of J0945+1009 (orange; \citealt{Plotkin15}), respectively; grey shaded regions refer to four different-ion BALs and the red shaded region highlights the newly emerged \civ\ BAL. The blue, green, red, and black dashed vertical lines locate at 2800, 1860, 1550, and 1400 \AA, respectively. Right panels: profiles of the four different-ion BALs (grey: \siiv, red: \civ, green: \aliii, blue: \mgii) normalized by the the local continuum fits in velocity space; dashed magenta horizontal lines represent the continuum level. The simultaneous \civ-BAL disappearance/emergence occurs after MJD = 55513. The inset panel shows the spectral fit of the \mgii\ BEL at MJD = 55513 (the HET spectrum has no coverage at $\lambda_{\rm rest}>2300$ \AA), in which the magenta line represents the best-fit of the \mgii\ BEL using the same prescription as  \citet{Shen11}, including a power-law continuum (dashed magenta), \feii\ emission (dotted blue), and the \mgii\ BEL (green). } 
      \label{J0827all_spec_fit}
\end{figure*}

The barely visible \civ\ BEL and weak \ciii]+\mgii\ emission features in the different spectroscopic epochs shown in Figure \ref{J0827all_spec_fit} indicate the WLQ nature of this LoBAL quasar. 
The four different-ion BALs at the same velocity have the same velocity width, suggesting that they arise in the same absorbing material. 
Note that the \siiv\ troughs are located at the blue edge of the spectra and are associated with large uncertainties; thus the detection of \siiv\ BALs is not as reliable as that of the other BALs.

\subsection{Continuum fits and EW measurements } \label{ew_measurement}
We adopt a reddened power-law model to fit the global continuum using the SMC-like reddening function from \citet{Pei92}. This model was fit to regions of the spectra that were visually identified to be relatively free of emission and absorption features (1260--1350 \AA, 1750--1800 \AA, 2100--2250 \AA, 3000--3100 \AA). 
Spectral slope and reddening are then derived from the global fits. Noticing that the WLQ nature of this LoBAL quasar, we choose another WLQ, J0945+1009 at $z=1.683$ from \citet{Plotkin15}, as a template to fit our spectra at different epochs by  optimizing the SMC-like reddening and scaling factors. 
As a comparison, we also consider a model combining a non-BAL composite spectrum \citep{VandenBerk01} as a template in combination with the same SMC-like reddening function to fit these spectra. 
The left panels of Figure \ref{J0827all_spec_fit} display all spectra of J0827+4252 obtained from SDSS and HET/LRS2 (adopting $z=2.038\pm0.005$ from \citealt{Hewett10}). 
There is clear trend that the weak emission-line feature becomes increasingly evident from low- to high-ionization emission lines, as noted by \citet{Plotkin15}; in addition, the \civ\ emission-line blueshift of J0827+4252 appears to be consistent with that of J0945+1009 ($\sim$5500 \kms). However, we require near-IR spectroscopy to establish the \civ\ emission-line blueshift, as J0945+1009 does show dramatic \mgii\ emission-line blueshift compared to the systemic redshift determined by the H$\beta$ emission line. As the global continuum fit may not be optimal for quantifying BAL properties, we choose the local continuum fit over these line-free regions using the same reddened power-law model for quantitative measurements. The local fits are in good agreement with global fits except for the spectrum at MJD = 57063 with an apparent change in spectral shape, possibly due to flux-calibration errors. 
To obtain BAL-disappearance timescales over different ions, we do not consider the highly uncertain spectrum at MJD = 54524 (S/N$\sim$3 in the continuum). 
Using a 2$\sigma$ EW measurement as the threshold of BAL-trough detection, \mgii\ BAL disappeared by MJD = 55513, \siiv\ and \aliii\ BALs disappeared by MJD = 57063, and the low-velocity \civ\ BAL disappeared by MJD = 58212. 
Our measurements of the spectra and BAL properties are tabulated in Table \ref{table2}. 
The time evolution of the EWs of these BALs is shown in the bottom panel of Figure \ref{J0827ew_mjd_seq}.

\begin{table}[h]
\centering
 \caption{ Measurements at different spectroscopic epochs}
 \begin{tabular}{cccccc}
  \hline\noalign{\smallskip}
 MJD & 52266 & 54524 & 55513 & 57063 & 58212 \\
 \hline
 $v_{\rm LOS}^b$ & $-$23.2  & $-$23.2 & $-$23.2 & $-$20.5 & $-$33 \\
$v_{\rm LOS}^r$ & $-$17.8  & $-$17.8 & $-$17.8 & $-$18.8 & $-$22 \\
$\alpha_{\lambda}$ & $-$1.87  & $-$1.1 & $-$2.2 & $-$4.97 & $-$1.32 \\
$r$ & 0.25  & 0.17 & 0.28 & 0.49 & 0.21 \\
\mgii\ & 6.3$\pm$1.01 & 3.2$\pm$1.15 & 0.6$\pm$0.57 & 0.2$\pm$0.45 &  \\
\aliii\ & 7.5$\pm$0.65  & 4.3$\pm$0.82 & 1.8$\pm$0.43 & 0.1$\pm$0.28 & 0 \\
\siiv\ & 13.5$\pm$1.8  & 4.7$\pm$4.8 & 3$\pm$1.36 & 0$\pm$0.53 & 0 \\
\civ$_{l}$ & 13.2$\pm$1.1  & 9.7$\pm$1.75 & 8.9$\pm$0.76 & 1.7$\pm$0.76 & 0 \\
\civ$_{h}$ & 0  & 0 & 0 & 1.2$\pm$0.7 & 26$\pm$0.7 \\
  \noalign{\smallskip}\hline
\end{tabular}
\label{table2}
\tablecomments{$v_{\rm LOS}^b$ and $v_{\rm LOS}^r$ are the blue- and red-edge velocities (1000 \kms) of each BAL trough. $\alpha_{\lambda}$ and $r$ are the index and reddening derived from the global reddened power-law fits. The last five  rows are the corresponding BAL EWs (\AA) over multiple ions at different epochs based on the local continuum fits, in which \civ$_{l}$ and \civ$_{h}$ represent the low-velocity and high-velocity \civ\ BALs, respectively. }
\end{table}

One conspicuous feature in Figure \ref{J0827ew_mjd_seq} is that the low-velocity \civ\ BAL slowly weakens until it completely disappears (over $>5$ yr in the rest frame), while a strong \civ\ BAL emerges abruptly (over $<1$ yr in the rest frame) at a higher velocity. Another dramatic feature is the faster disappearance of LoBALs compared to HiBALs. 
These observational results provide diagnostics for investigating the dominant driver of the BAL disappearance/emergence and the structure of outflows (see Section \ref{main_driver}). 

\subsection{Black-hole mass estimate}
\label{BH_mass}

Based on the single-epoch virial relation, \citet{Shen11} estimated the \mgii-based black hole (BH)  mass and Eddington ratio for this quasar to be $\sim6.3\times 10^{9}M_{\odot}$ and $\lambda_{\rm Edd}\sim$ 0.01. 
Note that the \mgii-based BH mass from \citet{Shen11} was measured from the spectrum at MJD = 52266 that has low S/N. We measured the \mgii\ BEL with a higher S/N spectrum at MJD = 55513 (see the inset panel of Figure~\ref{J0827all_spec_fit}) using the same prescription from \citet{Shen11}, which yields a FWHM$_{\rm {MgII}}=6990\pm610$ \kms\ after subtracting the power-law continuum and \feii\ emission (see the dashed and dotted lines in the inset panel of Figure \ref{J0827all_spec_fit}). Our measurement of the FWHM is consistent with that of \citet{Shen11}, where they measured FWHM$_{\rm {MgII}}=8860\pm1340$ \kms. \break 
The continuum fits indicate an approximately equal luminosity at 3000~\AA\ over the two epochs for this quasar (so log$M_{\rm BH} \propto $ logFWHM$^2$). Our measurement yields $M_{\rm BH}\sim 4\times 10^{9}M_{\odot}$.

We caution that the uncertainties on the BH mass not only include measurement errors ($\sim$0.1 dex) and systematic errors ($\sim$0.3--0.5 dex) of the single-epoch scaling approach, but also may have additional systematic errors related to non-virial gas motions sometimes seen in WLQs (e.g., \citealp{Plotkin15}).

\section{ Discussion} \label{main_driver}

\begin{figure}[h]
\center{}
 \includegraphics[height=7cm,width=8.5cm,  angle=0]{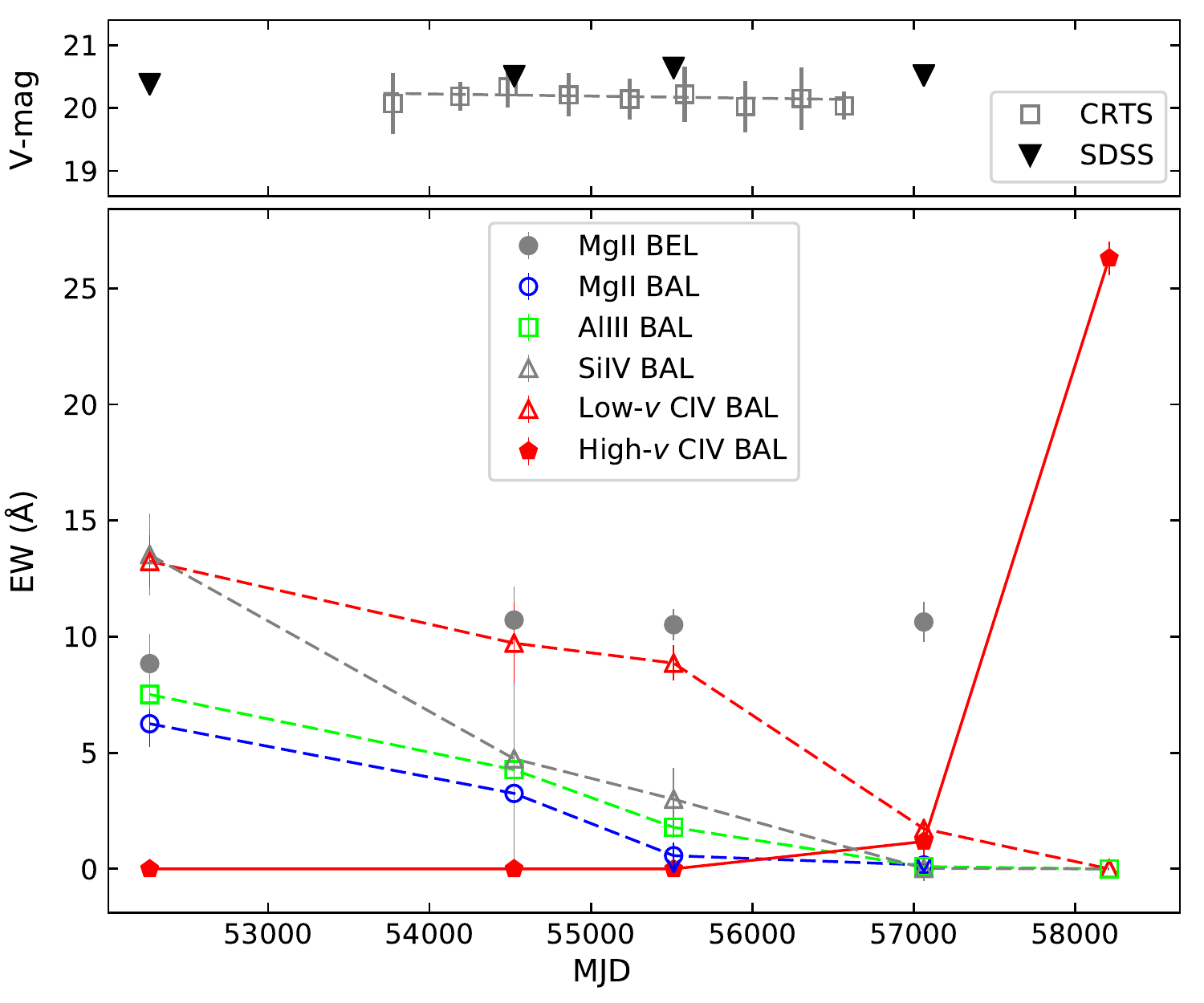} 
      \caption{ Top panel: the synthetic \textit{V}-band light curve obtained from CRTS, in which grey squares are mean values sampled within one-year intervals (the error bar is determined from the standard deviation in a corresponding interval). Inverted black triangles are the synthetic \textit{V}-band magnitudes from different-epoch SDSS spectra. The dashed grey line represents the best linear fit. Bottom panel: MJD versus BAL/BEL EW, in which \civ\ (red triangles), \siiv\ (grey triangles), \aliii\ (green squares), and \mgii\ (blue circles) BALs are indicated. Downward arrows represent upper limits on EWs. The absolute \mgii-BEL EWs are marked as filled grey dots. Red pentagons represent the newly-emerged \civ\ BAL at a higher velocity compared to the disappearing \civ\ BAL. }
      \label{J0827ew_mjd_seq}
\end{figure}
Ionization changes of BAL absorbers and transverse motions across our LOS are the two most widely accepted explanations accounting for BAL variability. We discuss these two possibilities in this section.

\subsection{ Ionization-change scenario} \label{ionization_change}

Assuming solar abundances under an optically thin condition, the observed BAL disappearance over multiple ions can be explained by ionic column density variations in response to ionization changes. 
Depending on the initial value of the ionization parameter, a reduction in the column density of an ion can be caused by either an increase or a decrease in the ionizing flux. From Figure~4 of \citet{Hamann01}, the ionization parameter at MJD = 52266 can be constrained to be log $U\sim-3$ due to the simultaneous presence of \mgii, \aliii, \civ, and \siiv\ BALs. The subsequent disappearance of \mgii\ and \aliii\ BAL troughs together with the weakening of \civ\ and \siiv\ BALs by MJD = 55513 can be explained by an increase in the ionizing flux (to log $U\sim-1.5$). 
The observations do not support a reduction in the ionizing flux as the \civ\ BAL would have disappeared before the \mgii\ BAL in such a scenario. 
A continued increase in the ionizing flux would result in the further weakening and eventual disappearance of the low-velocity \civ\ BAL. 
However, the simultaneous disappearance of the low-velocity \civ\ BAL and appearance of the high-velocity \civ\ BAL at MJD = 58212  poses some problems. It is tempting to argue that the increase in the ionizing flux resulted in favorable ionization conditions for the \civ\ BAL to arise in high-velocity material  at a larger distance. If the source of the ionizing flux is in common for the two absorbers, the fractional changes in the ionization parameter should be comparable for the two absorbers. However, the fractional ionization-parameter change required to explain the appearance of the high-velocity \civ\ BAL is much larger than that needed to explain the disappearance of the low-velocity \civ\ BAL.

The complete disappearance of the moderately strong low-velocity \civ\ BAL clearly requires a large change in the ionizing flux. 
Note that saturation in the absorption troughs (supported by the coexistence of \siiv, \civ, \aliii, and \mgii\ BALs at the same velocity) demands an even larger change in the ionizing flux. 
Therefore, we likely expect to see variations in the UV light curve even though the actual ionizing photons are in the far-UV. To assess the UV continuum variability, we analyzed the synthetic \textit{V}-band light curve (corresponding to the UV band in the rest frame) obtained from the Catalina Real-Time Transient Survey (CRTS; \citealp{Drake09}), as depicted in the top panel of Figure~\ref{J0827ew_mjd_seq}. 
As we are interested in the long-term continuum variability, we binned the light curve in one-year intervals (the  grey squares). 
As a comparison, we also show the upper limit (due to possible flux missing from the spectroscopic aperture) on the \textit{V}-band magnitudes  (inverted black triangles in the top panel of Figure \ref{J0827ew_mjd_seq}) derived from the SDSS spectra using the prescription  from \citet{Jester05}. 
Both of the $V$-band light curves do not point to any significant changes in the continuum flux. 
Additionally, the EW measurements of the \mgii\ BEL ($2780<\lambda_{\rm rest}<2820$ \AA) indicate that the \mgii\ BEL did not experience substantial variations (see filled grey dots in the bottom panel of Figure~\ref{J0827ew_mjd_seq}) during our observations. 

We also explored the radiation pressure confinement (RPC) model of \citet{Baskin2014a}, which proposes a radial density profile, leading to a range of $U$ in a single absorber system. Given the exponentially increasing density profile (equation 22 of \citealt{Baskin2014a}), the high density, low-$U$ gas (where the \mgii\ ions are formed) should be less sensitive to changes in the ionizing flux as compared to the low-density, high-$U$ gas (where the \civ\ ions are formed). Therefore, ionization changes in the context of the RPC model are unlikely to explain our observation of faster disappearance of LoBAL troughs. 

In light of the observational evidence discussed above, we conclude that the ionization-change mechanism is not a dominant driver of the BAL disappearance/emergence, although it may play a role in a single ``direction'' (BAL disappearance or BAL emergence).

\subsection{Absorbing-gas motion scenario} \label{transverse_motion}

We first assess the possibility of an accelerating BAL absorber moving from low to high velocity along our LOS. We did not find any monolithic shifts of the \civ\ BAL before its disappearance, and there are no \aliii\ BAL (see Figure \ref{J0827all_spec_fit}) appearing at a similar velocity as the newly emerged \civ\ BAL. 
Additionally, if the emerging \civ\ BAL were caused by the acceleration of the disappearing BAL, then the derived acceleration ($\approx$ 25 cm s$^{-2}$) would be about two orders-of-magnitude larger than the typical upper limit of BAL acceleration found in previous studies (e.g., \citealp{Grier16}). 
Therefore, the acceleration of outflows along our LOS can be ruled out.

Next, we examine the case where BAL material moves across our LOS. 
In reality, a BAL absorber system may be composed of numerous ``clouds'' with a small volume filling factor (e.g., \citealp{Hamann13}), and the BAL-absorber density, as well as the background source emission,  are most likely inhomogeneous. For simplicity, we consider a single cloud with a uniform density along our LOS. The BAL disappearance/emergence can be explained by a scenario in which one gas cloud containing both HiBAL/LoBAL absorbers takes more than five years to move out of our LOS; then, another cloud producing only \civ\ absorption spends less than one year moving into our LOS. 
The faster disappearance of the LoBALs can be explained if the low-ionization gas is embedded inside the high-ionization gas and has a smaller covering factor (e.g., \citealp{Arav99,Baskin2014a}). Observationally, such an explanation is supported by a recent study from \citet{Hamann18}, where they found that the low-ionization/high-column gas tend to have smaller covering fractions in BAL outflows based on a large BALQSO sample. 
No \aliii\ BAL appears at a similar velocity to the emerging \civ\ BAL in the HET spectrum. This could be due to either the low-ionization gas not yet moving into our LOS or the material associated with the emerging \civ\ BAL having a relatively high level of ionization.

To estimate the continuum source size, we choose the \mgii-based BH mass (4$\times10^9 M_{\odot}$) of this quasar (see Section \ref{BH_mass}). Following \citet{Rogerson16}, the lower limit on the projected size of the continuum region is estimated to be $\sim$ 0.01 pc. 
The average trough depth ($\sim$0.6) of the disappearing \civ\ BAL indicates that the BAL absorber covers at least 60\% of the projected area of the continuum source. 
To set a lower limit on the transverse velocity (see  \citealt{Rogerson16} for details), we consider the homogeneous, sharp-edged case where $D_{t} = f\times D_{b}$ ($D_{t}$, $f$, and $D_{b}$ refer to the crossing distance, LOS covering factor, and  the projected size of the background source, respectively). The lower limit on the transverse velocity is estimated to be $v_{\rm t}>$ 1100 \kms\ using the BAL-disappearance timescale (5.36 yr). As a comparison, the mean radial LOS velocity of the disappearing \civ\ BAL is 20000 \kms.  If we assume that the transverse velocity is comparable to the Keplerian velocity (see, e.g., \citealp{McGraw2017}), then the derived distance between the absorber and the central engine is 14.2 pc.

For the newly emerged \civ\ BAL at $v_{\rm LOS}\sim 0.1c$, a lower limit on the transverse velocity is derived to be 5870 \kms\ using the change of trough depth ($\sim$0.6) and BAL emergence timescale ($<1$ yr). Similarly, the derived distance between the absorber and the central engine is less than 0.5 pc assuming the Keplerian velocity. We require follow-up spectroscopy to trace a time evolution (at least the whole emerging process) for the newly emerged BAL trough and obtain a better estimate of BH mass which, in turn, can be used to derive an upper limit on the transverse velocity and constrain its ionization level as well as  outflow structure to the low-velocity BAL absorber.

\section{Summary and future work}\label{disc_con}

We report a case of BAL disappearance/emergence in a WLQ for the first time, where LoBALs at the same velocity disappeared faster than HiBALs, followed by a quick emergence of a strong \civ\ BAL at a higher velocity.  
Based on the simultaneous BAL disappearance/emergence and non-variability of the continuum flux, we disfavor a scenario where an ionization change in the absorbing material resulted in the observed BAL variability. Transverse motion of the absorption system is the most likely dominant driver of  BAL disappearance/emergence for this quasar. 

Such dramatic BAL disappearance over multiple ions, the quick emergence of a strong BAL trough, and the presence of mildly relativistic outflows in this WLQ, together suggest an extreme, wind-dominated phase, in which BAL absorbers are expected to cross our LOS more quickly and frequently than in other BAL quasars. 
Multi-wavelength follow-up observations will be helpful for further investigations regarding physical properties of the WLQ, the connection between WLQs and BALQSOs, the relatively roles of BAL/BEL outflows and their potential in expelling material from inner to large-scale regions. 
Particularly, UV/optical/near-IR follow-up spectroscopy will allow us to establish the \civ-BEL blueshift, trace subsequent BAL/BEL variability (possible LoBAL emergence), and obtain a better estimate of the BH mass, which is critical for probing the physical nature of this LoBAL WLQ.

\acknowledgments
We thank Michael Eracleous, Xiaohui Fan, and Patrick Hall for stimulating discussions. We thank Richard Plotkin for kindly providing the WLQ spectra of J0945+1009 obtained by VLT/X-shooter. 
We are grateful to Robin Ciardullo and Michael Eracleous for assistance with the observations by the Hobby-Eberly Telescope. We thank Xiaohui Fan, Jonathan Trump, Feige Wang, and Jinyi Yang for help in preparing follow-up observations. 

W. Yi thanks financial support from the China Scholarships Council (No. 201604910001) for his postdoctoral study at the Pennsylvania State University. 
W. Yi also thanks support from the Chinese National Science Foundation (NSFC-11703076) and the West Light Foundation of The Chinese Academy of Sciences (Y6XB016001). MV and WNB acknowledge support from NSF grant AST-1516784. JT acknowledges support from NASA ADP grant 80NSSC18K0878. 
NFA thanks TUBITAK (115F037) for financial support.  F.V. acknowledges financial support from CONICYT and CASSACA through the Fourth call for tenders of the CAS-CONICYT Fund. The Cosmic Dawn Center is funded by the DNRF. 

Funding for SDSS-III has been provided by the Alfred P. Sloan Foundation, the Participating Institutions, the National Science Foundation, and the U.S. Department of Energy Office of Science. 
The Low-Resolution Spectrograph 2 (LRS2) was developed and funded by the
University of Texas at Austin McDonald Observatory and Department of
Astronomy, and by the Pennsylvania State University. We thank the
Leibniz-Institut f\"ur Astrophysik Potsdam and the Institut f\"ur
Astrophysik G\"ottingen for their contributions to the construction
of the integral field units.

The Hobby-Eberly Telescope (HET) is a joint project of the University of Texas at Austin, the Pennsylvania State University, Ludwig-Maximillians-Universit\"{a}t M\"{u}nchen, and Georg-August-Universit\"{a}t G\"{o}ttingen. The Hobby-Eberly Telescope is named in honour of its principal benefactors, William P. Hobby and Robert E. Eberly.

\newpage

\end{document}